\documentclass[twocolumn,showpacs,amsmath,amssymb,superscriptadress,prc]{revtex4-1}
\usepackage{epsfig,amsmath,graphicx,amssymb,tabularx}
\usepackage[dvipdfmx,bookmarks=true,colorlinks,citecolor=blue,linkcolor=blue,anchorcolor=blue,filecolor=blue,urlcolor=blue]{hyperref}
\usepackage{CJK}
\usepackage{dcolumn}
\usepackage{bm}
\usepackage{multirow}

\def\beq{\begin{equation}}
\def\eeq{\end{equation}}
\def\bea{\begin{eqnarray}}
\def\eea{\end{eqnarray}}
\begin{document}
\begin{CJK*}{GBK}{}

\title{The Role of Tensor Force in Heavy-Ion Fusion Dynamics}


\author{Lu Guo$^{1}$}
\email{luguo@ucas.ac.cn}
\author{C\'edric Simenel$^{2}$}
\email{cedric.simenel@anu.edu.au}
\author{Long Shi$^{1}$}
\author{Chong Yu$^{1}$}

\affiliation{$^1$ School of Physics, University of Chinese Academy of Sciences, Beijing 100049, China}
\affiliation{$^2$ Department of Nuclear Physics, RSPE, Australian National University, Canberra, Australia}

\date{\today}

\begin{abstract}
The tensor force is implemented into the time-dependent Hartree-Fock (TDHF) theory so that both exotic and stable
collision partners, as well as their dynamics in heavy-ion fusion, can be described microscopically.
The role of tensor force on fusion dynamics is systematically investigated for
$^{40}\mathrm{Ca}+\mathrm{^{40}Ca}$, $^{40}\mathrm{Ca}+\mathrm{^{48}Ca}$, $^{48}\mathrm{Ca}+\mathrm{^{48}Ca}$,
$^{48}\mathrm{Ca}+\mathrm{^{56}Ni}$, and $^{56}\mathrm{Ni}+\mathrm{^{56}Ni}$ reactions
which vary by the total number of spin-unsaturated magic numbers in target and projectile.
A notable effect on fusion barriers and cross sections
is observed by the inclusion of tensor force.
The origin of this effect is analyzed.
The influence of isoscalar and isovector tensor terms is investigated with the T$IJ$ forces.
These effects of tensor force in fusion dynamics are essentially attributed to the shift of low-lying vibration states of
colliding partners and nucleon transfer in the asymmetric reactions.
Our calculations of above-barrier fusion cross sections also show that tensor force does not significantly affect the
dynamical dissipation at near-barrier energies.
\end{abstract}


\maketitle
\end{CJK*}

The tensor interaction is of great interests in nuclear physics.
It is indeed crucial to explain the properties of the deuteron. 
It plays a significant role in nuclear structure, in particular
in exotic nuclei far from stability with extreme ratios between proton number and neutron number. 
The fact that exotic nuclei
may show distinct characteristics from those seen in usual stable nuclei
could in part be attributed to tensor interaction. In particular,
in nuclear dynamics, the tensor force changes not only the spin-orbit splitting, but also the intrinsic excitations
which may give rise to dynamical effects and are more complicated than those arising from simple shell evolution.
The introduction of tensor force
improved the systematic agreement between model predictions and experimental data in the shell evolution of exotic
nuclei~\cite{Otsuka2006_PRL97-162501}, spin-orbit splitting~\cite{Colo2007_PLB646-227},
Gamow-Teller and charge exchange spin-dipole excitations~\cite{Bai2010_PRL105-072501}.
In spite of these indications in nuclear structure,
most calculations for reaction dynamics ignored tensor force for decades both in macroscopic-microscopic approaches and self-consistent mean-field methods. 

The microscopic mechanisms involved in low-energy
heavy-ion collisions originate from the effective interaction between the nucleons,
the understanding of which remains one of the main challenges in nuclear physics.
Microscopic models of heavy-ion collisions~\cite{Negele1982_RMP54-913,Wen2013_PRL111-012501,Wen2014_PRC90-054613}
can then be used to test some properties of the nuclear force. 
In such collisions, various couplings between
the relative motion and the internal degrees of freedom of colliding partners complicate the description of the reaction mechanisms.
Through coupling, these internal degrees of freedom, which include low-lying vibrations~\cite{Morton1994_PRL72-4074,Stefanini1995_PRL74-864,Simenel2013_PRC88-064604},
nucleon transfer~\cite{Washiyama2009_PRC80-031602,Simenel2010_PRL105-192701,Sekizawa2017_PRC96-041601}, rotations~\cite{Leigh1995_PRC52-3151}, and
high-lying giant resonances~\cite{Diaz-Torres2008_PRC78-064604} have been shown to
modify the reaction dynamics and, subsequently, the outcome of the
reaction itself~\cite{Dasso1985_NPA432-495}.

The effect of tensor force on these couplings in heavy-ion fusion dynamics is an open question which we address in the present study.
In particular, low-lying collective states and nucleon transfer, which have the strongest impact on near-barrier fusion,
are very sensitive to the underlying shell structure, which is in turn affected by the tensor force.
Another possible effect of tensor force is to modify dissipation in heavy-ion collisions.
The latter has been the subject of theoretical studies at energies well above the Coulomb barrier~\cite{Dai2014_SciChinaPMA57-1618,
Stevenson2016_PRC93-054617,Shi2017_NPR34-41}.
To our knowledge, however, the interplay between the tensor force and dissipation at near-barrier energies has never been investigated.
Our work is the first attempt at a study of the effects induced by tensor force
on the low-energy fusion dynamics.

In order to study the role of tensor force in fusion dynamics, it is desirable to develop a theoretical framework in which the dynamical effects,
such as couplings and dissipation, are all treated on the same footing.
To this end, we incorporated the full tensor terms of Skyrme energy density functional (EDF) into the microscopic
time-dependent Hartree-Fock (TDHF) approach,
which automatically includes one-body dissipation as well as the couplings between collective motion
and internal degrees of freedom at the mean-field level.
TDHF theory is a well-defined microscopic framework and extensively applied to the study of
heavy-ion collisions~\cite{Umar2006_PRC74-021601,Guo2007_PRC76-014601,Guo2008_PRC77-041301,
Simenel2011_PRL106-112502,Guo2012_EPJWoC38-09003,Sekizawa2013_PRC88-014614,
Umar2014_PRC89-034611,Vophuoc2016_PRC94-024612,Wang2016_PLB760-236,Umar2017_PRC96-024625,Yu2017_SciChinaPMA60-092011,Williams2018_PRL120-022501}
and vibration dynamics~\cite{Lacroix2004_PPNP52-497,Maruhn2005_PRC71-064328,Nakatsukasa2005_PRC71-024301,
Umar2005_PRC71-034314,Reinhard2007_EPJA32-19,Simenel2009_PRC80-064309,Fracasso2012_PRC86-044303,Avez2013_EPJA49-76}.
For recent reviews, see Refs.~\cite{Simenel2012_EPJA48-152,Nakatsukasa2016_RMP88-045004}.

Most TDHF calculations employ Skyrme effective interaction~\cite{Skyrme1956_PM1-1043},
in which the two-body tensor force was proposed in its original form as
\begin{align}
\begin{split}
v_T&=\dfrac{t_\mathrm{e}}{2}\bigg\{\big[3({\sigma}_\mathrm{1}\cdot\mathbf{k}')({\sigma}_\mathrm{2}\cdot\mathbf{k}')-({\sigma}_\mathrm{1}\cdot{\sigma}_\mathrm{2})\mathbf{k}'^{\mathrm{2}}\big]\delta(\mathbf{r}_\mathrm{1}-\mathbf{r}_\mathrm{2})\\
&+\delta(\mathbf{r}_\mathrm{1}-\mathbf{r}_\mathrm{2})\big[3({\sigma}_\mathrm{1}\cdot\mathbf{k})({\sigma}_\mathrm{2}\cdot\mathbf{k})-({\sigma}_\mathrm{1}\cdot{\sigma}_\mathrm{2})\mathbf{k}^\mathrm{2}\big]\bigg\}\\
&+t_\mathrm{o}\bigg\{3({\sigma}_\mathrm{1}\cdot\mathbf{k}')\delta(\mathbf{r}_\mathrm{1}-\mathbf{r}_\mathrm{2})({\sigma}_\mathrm{2}\cdot\mathbf{k})-({\sigma}_\mathrm{1}\cdot{\sigma}_\mathrm{2})\mathbf{k}'
\delta(\mathbf{r}_\mathrm{1}-\mathbf{r}_\mathrm{2})\mathbf{k}\bigg\}.
\end{split}
\end{align}
The coupling constants $t_\textrm{e}$ and $t_\textrm{o}$ represent the strengths of triplet-even and
triplet-odd tensor interactions, respectively.  The operator $\mathbf{k}=\frac{1}{2i}(\nabla_1-\nabla_2)$ acts on the right, and
$\mathbf{k}'=-\frac{1}{2i}(\nabla'_1-\nabla'_2)$ acts on the left.

With the inclusion of central, spin-orbit and tensor forces, the full version of Skyrme EDF is expressed as
\begin{align}
\label{EDFH}
\begin{split}
\mathcal{H}&=\mathcal{H}_0+\sum_{\rm{t=0,1}}\Big\{C_{\rm{t}}^{\rm{s}}\mathbf{s}_{\rm{t}}^2+C_{\rm{t}}^{\Delta{s}}
\mathbf{s}_{\rm{t}}\cdot\Delta\mathbf{s}_{\rm{t}}+C_{\rm{t}}^{\nabla s}(\nabla\cdot \mathbf{s}_{\rm{t}})^2 \\
&+C_{\rm{t}}^{F}\big(\mathbf{s}_{\rm{t}}\cdot
\mathbf {F}_{\rm{t}}-\frac{1}{2}\big(\sum_{\mu=x}^{z}J_{\rm{t}, \mu\mu}\big)^2-\frac{1}{2}\sum_{\mu, \nu=x}^{z}J_{\rm{t}, \mu\nu}J_{\rm{t}, \nu \mu}\big)\\
&+C_{\rm{t}}^{\rm{T}}\big(\mathbf{s}_{\rm{t}}\cdot\mathbf{T}_{\rm{t}}-
\sum_{\mu,\nu=x}^{z}J_{\rm{t},\mu\nu}J_{\rm{t},\mu\nu}\big)\Big\},
\end{split}
\end{align}
where $\rm{t}$ denotes the isospin and $\mathcal{H}_0$ is the simplified functional without tensor terms as used in Sky3D code~\cite{Maruhn2014_CPC185-2195} and most TDHF calculations.
The coupling constants $C$ have been defined in Refs.~\cite{Lesinski2007_PRC76-014312,Davesne2009_PRC80-024314}.
Due to the computational complexity, various approximations to nuclear force have been employed, which restrict the number of degrees of freedom accessible during a collision, and hence the nature and degree of
dissipation dynamics.
For instance, the inclusion of spin-orbit interaction solved the conflict between TDHF predictions and experiments~\cite{Umar1986_PRL56-2793},
and turned out to play an important role in fusion and dissipation dynamics~\cite{Maruhn2006_PRC74-027601,Dai2014_PRC90-044609}.
In our code, we incorporated all the time-even and time-odd density dependence of the tensor terms shown in Eq.~(\ref{EDFH}).
Our TDHF calculations have been performed in a fully three-dimensional Cartesian space and without any symmetry restriction.
As pointed out in Refs.~\cite{Lesinski2007_PRC76-014312,Stevenson2016_PRC93-054617}, the terms containing the gradient of spin density may cause the spin instability in both nuclear structure and reaction studies, so we set $C_{\rm{t}}^{\Delta{s}}=C_{\rm{t}}^{\nabla s}=0$ in our calculations.

The Skyrme tensor force has been constructed in two ways. One is to add perturbatively to the existing standard interactions, for instance,
the existing Skyrme parameter SLy5~\cite{Chabanat1998_NPA635-231} plus tensor force, denoted as SLy5t~\cite{Colo2007_PLB646-227}.
The comparison between calculations with SLy5 and SLy5t addresses the question on how much of the changes is caused by tensor force itself.
Another approach is to readjust the full set of Skyrme parameters self-consistently.
This strategy has been adopted in Ref.~\cite{Lesinski2007_PRC76-014312} and led to
the set of T$IJ$ parametrizations with a wide range of isoscalar and isovector tensor couplings.
Due to its fitting strategy, the contributions from the tensor force and the rearrangement of all other terms could be physically entangled.

The contribution of tensor force is expected to be nearly zero for the ground state of spin-saturated nuclei,
whereas it might be significant for the nuclei with
one level out of two spin-orbit partners filled. We have thus chosen five representative reactions $^{40}\mathrm{Ca}+\mathrm{^{40}Ca}$,
$^{40}\mathrm{Ca}+\mathrm{^{48}Ca}$, $^{48}\mathrm{Ca}+\mathrm{^{48}Ca}$, $^{48}\mathrm{Ca}+\mathrm{^{56}Ni}$, and $^{56}\mathrm{Ni}+\mathrm{^{56}Ni}$
quantified by $N_s=0,1,2,3,$ and 4, respectively, where $N_s$ is the total number of spin-unsaturated magic number in target and projectile.
In these collisions, the reaction partners are closed-shell corresponding to 20 (spin-saturated) and 28 (spin-unsaturated) neutron or proton magic numbers.

\begin{figure}
\includegraphics[width=7 cm]{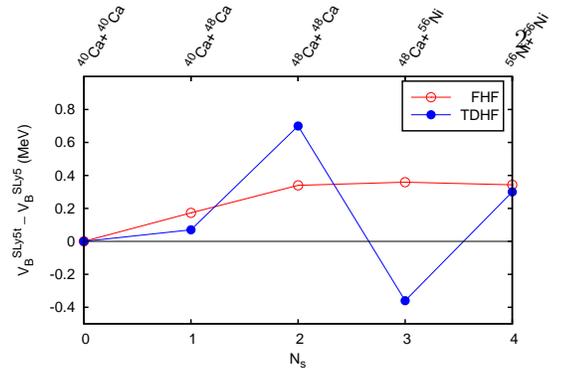}
\caption{(Color online.) Difference of fusion barrier heights with tensor (SLy5t) and without tensor (SLy5) computed with the FHF static (open circles) and TDHF dynamical (solid circles) methods.
\label{Fig:barriers}}
\end{figure}

The effect of tensor force on fusion could have static (e.g., modification of ground-state density)
and dynamical (e.g., modification of couplings and dissipation) origins.
In order to disentangle the contributions from static and dynamical effects, both frozen Hartree-Fock (FHF), where the nuclei keep their ground-state densities \cite{Brueckner1968_PR173-944,Simenel2008_IJMPE17-31,Washiyama2008_PRC78-024610,Simenel2013_PRC88-064604,Shi2017_NPR34-41,Simenel2017_PRC95-031601},
and TDHF calculations of fusion barriers have been performed. Dynamical fusion barriers can be computed directly from TDHF \cite{Simenel2008_IJMPE17-31,Washiyama2008_PRC78-024610,Shi2017_NPR34-41}, or with the density-constrained TDHF technique \cite{Umar2006_PRC74-021601}.

Figure~\ref{Fig:barriers} shows the variation of static (FHF) and dynamic (TDHF) fusion barrier heights due to the tensor force in SLy5t as a function of $N_s$.
For the spin-saturated reaction $^{40}\mathrm{Ca}+\mathrm{^{40}Ca}$, the static barrier obtained from FHF calculations with SLy5t is same as with SLy5, as expected, due to the nearly zero contributions of tensor force to the ground state of spin-saturated nucleus $^{40}\mathrm{Ca}$.
For the other spin-unsaturated reactions, the static (FHF) barrier with SLy5t is systematically higher than SLy5.
This indicates a fusion hindrance due to tensor force in this mass region.

The variation of dynamical barriers due to tensor force presents different behavior from the static results,
as shown in Fig.~\ref{Fig:barriers}.
The dynamical barriers appear staggering as a function of $N_s$.
In particular, a distinguishing characteristics is the lower dynamical barrier by the inclusion of tensor force in $^{48}\mathrm{Ca}+\mathrm{^{56}Ni}$. This indicates the importance of dynamical effects which can remarkably modify the barrier height.
These dynamical effects, which could be low-lying vibration states, nucleon transfer, and dynamical dissipation,
strongly affect the near-barrier fusion. In the following, we will investigate these three dynamical effects and their interplay
in the role of tensor force on fusion dynamics. For symmetric reactions, the role of nucleon transfer
is minimized. Therefore, the effect of tensor force on the vibrational modes will be first studied focusing on symmetric reactions.
In a second step, the role of tensor force in nucleon transfer will be investigated in asymmetric reactions.

\begin{figure}
\flushleft
\includegraphics[width=8 cm]{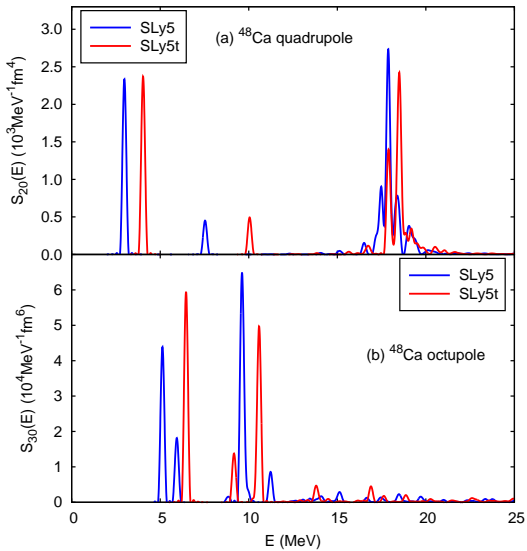}
\caption{(Color online.) Strength function of the quadrupole (a) and octupole (b) excitations in $^{48}\mathrm{Ca}$.}
\label{vibration-Ca48}
\end{figure}

The effect of tensor terms on the isovector giant dipole resonance has been studied in Ref.~\cite{Fracasso2012_PRC86-044303}.
Here, we focus on the low-lying quadrupole and octupole modes which have a stronger impact on the near-barrier fusion.
The peak energies of low-lying vibration states, RPA results~\cite{Cao2009_PRC80-064304}, and
experimental data~\cite{Raman1987_ADNDT36-1,Kibedi2002_ADNDT80-35} are listed in Tab~\ref{tab:Peak}.
In the quadrupole excitation of $^{40}\mathrm{Ca}$, we did not find the low-lying collective states.
This is consistent with the fact that, in this nucleus, the $2_1^+$ state
does not exhibit a strong increase of collectivity as compared to the single-particle picture~\cite{Raman1987_ADNDT36-1}.
The $3_1^-$ state of octupole excitation in $^{40}\mathrm{Ca}$ is pushed 0.13 MeV down  by tensor force.
This indicates the tensor force has small effect on the vibration states of $^{40}\mathrm{Ca}$ and, subsequently,
the fusion dynamics through the couplings.
This is consistent with the barrier in $^{40}\mathrm{Ca}$+$^{40}\mathrm{Ca}$ as shown in Fig.~\ref{Fig:barriers}.
The other states are at higher energies and have a smaller strength.
This $3_1^-$ state is then expected to have a much stronger effect on the near-barrier reaction mechanisms than the other octupole states.
We also observe the difference in the peak energies of vibration states between TDHF and RPA approaches, although the trend of energy variance
induced by tensor force is the same. This could possibly
arise from the different treatment of center-of-mass (c.m.) corrections.
Since c.m. corrections lead to ambiguities in TDHF calculations of heavy-ion collisions, they are neglected in present vibration studies to allow a consistent treatment of structure and dynamics of colliding partners. The inclusion of c.m. corrections in vibration states and a detailed
comparison between TDHF and RPA calculations will be the subject of future works.

\begin{table}
\caption{Peak energies (in MeV) of low-lying vibration states in $^{40}\mathrm{Ca}$, $\mathrm{^{48}Ca}$, and $\mathrm{^{56}Ni}$ with and without tensor force.
The RPA results~\cite{Cao2009_PRC80-064304} and experimental data~\cite{Raman1987_ADNDT36-1,Kibedi2002_ADNDT80-35} are also shown.}
\label{tab:Peak}
\begin{center}
\begin{tabular*}{0.47\textwidth}{@{\extracolsep{\fill}}cccccc}
\hline\hline
Nucleus                             & State  & Method    &$E_\nu$(SLy5)&$E_\nu$(SLy5t)&  Expt.   \\ \hline
{$^{40}\mathrm{Ca}$} & $3_1^-$ &  This work   & 3.64     & 3.51     &  3.74    \\
                     &         &  RPA         & 3.78     & 3.02     &          \\
\hline
$^{48}\mathrm{Ca}$ &  $2_1^+$  &  This work   & 2.98     & 4.03     &  3.83    \\
                   &           &  RPA         & 3.05     & 3.88     &          \\
                   &  $3_1^-$  &  This work   & 5.13     & 6.46     &  4.51   \\
                   &           &  RPA         & 4.78     & 6.16     &         \\
\hline
$^{56}\mathrm{Ni}$ &  $2_1^+$  &  This work   & 2.30     &  2.76    &  2.70    \\
                   &  $3_1^-$  &  This work   & 9.53     &  9.68    &         \\
\hline \hline
\end{tabular*}
\end{center}
\end{table}

The situation is quite different for the spin-unsaturated nucleus $^{48}\mathrm{Ca}$. As seen in Fig.~\ref{vibration-Ca48},
the tensor force shifts both the quadrupole and octupole strength distributions to higher energies. In particular, the energy of $2_1^+$ ($3_1^-$)
state is pushed 1.05 MeV (1.33 MeV) up as shown in Tab.~\ref{tab:Peak}.
Fusion is dynamically hindered by tensor force in $^{48}\mathrm{Ca}$-involved reactions due to the increase of the energy of the low-lying collective vibrations.
This is a plausible explanation for the increase of dynamical barrier due to tensor force in the symmetric system $^{48}\mathrm{Ca}$+$^{48}\mathrm{Ca}$ ,
as shown in Fig.~\ref{Fig:barriers}. We also observe that the $2_1^+$ state by the inclusion of tensor force in SLy5t is
reasonably close to the experimental value of 3.83 MeV~\cite{Raman1987_ADNDT36-1}. However, the $3_1^-$ state
apparently overestimates the experiment~\cite{Kibedi2002_ADNDT80-35}, which may be due to the limitation of mean-field approximation and the choice of the interaction.

\begin{table}
\caption{Average number of transferred neutrons (n) and protons (p) with and without tensor force in  asymmetric
central collisions at 0.1~MeV below the TDHF barrier. The $+(-)$ sign indicates transfer from the light (heavy) fragment. }
\label{tab:transfer}
\begin{center}
\begin{tabular*}{0.47\textwidth}{@{\extracolsep{\fill}}ccccc}
\hline\hline
Reactions  &  $\mathrm{N_n}$(SLy5) &  $\mathrm{N_n}$(SLy5t) &   $\mathrm{N_p}$(SLy5) &   $\mathrm{N_p}$(SLy5t) \\ \hline
$^{40}\mathrm{Ca}$+$^{48}\mathrm{Ca}$ &  -0.85  &  -0.25   &  +0.95   &  +0.94    \\
$^{48}\mathrm{Ca}$+$^{56}\mathrm{Ni}$ &  +1.37  &  +1.88   &  -0.29   &  -1.06   \\
\hline \hline
\end{tabular*}
\end{center}
\end{table}

The similar effect of tensor force appears in the exotic nucleus $^{56}\mathrm{Ni}$, but the effect is much weaker than
in $^{48}\mathrm{Ca}$ due to the underlying shell structure.
This weaker effect results in a smaller increase of dynamical barrier in
$^{56}\mathrm{Ni}$+$^{56}\mathrm{Ni}$ compared with $^{48}\mathrm{Ca}$+$^{48}\mathrm{Ca}$, as shown in Fig~\ref{Fig:barriers}.

For the asymmetric reactions, both the low-lying vibrations and nucleon transfer are crucial for the fusion dynamics.
As the tensor force affects the single-particle levels, this could clearly affect transfer and then the TDHF fusion barrier.
We calculate the transferred nucleon numbers with and without tensor force at the sub-barrier collisions,
as listed in Tab.~\ref{tab:transfer}.
For $^{40}\mathrm{Ca}$+$^{48}\mathrm{Ca}$, less neutrons are transferred from $^{48}\mathrm{Ca}$ to $^{40}\mathrm{Ca}$
in SLy5t, and the transferred protons going from $^{40}\mathrm{Ca}$ to $^{48}\mathrm{Ca}$ are
nearly same in SLy5 and SLy5t. This small transfer (less than one neutron in average) results in a negligible effect on the fusion barrier.
As observed in Fig.~\ref{Fig:barriers}, the small variation of dynamical barrier in this reaction indicates the vibrational couplings are
dominated by the $3_1^-$ state in $^{40}\mathrm{Ca}$, which is slightly affected by tensor force. In addition, the fact that there is only
one spin-unsaturated magic number also accounts for the small effect of tensor force both in static and dynamic barriers.

The nucleon transfer presents distinct behavior in $^{48}\mathrm{Ca}$+$^{56}\mathrm{Ni}$. More neutrons are
transferred from $^{48}\mathrm{Ca}$ to $^{56}\mathrm{Ni}$ in SLy5t because the energy gap between the most high-lying
occupied orbit $1f_{7/2}$ in $^{48}\mathrm{Ca}$ and the most low-lying unoccupied orbit $2p_{3/2}$ in $^{56}\mathrm{Ni}$
is reduced by about 1 MeV by the tensor force. The large neutron transfer
decreases the dynamical barrier~\cite{Jiang2014_PRC89-051603}.
Also more protons are observed to transfer from $^{56}\mathrm{Ni}$ to $^{48}\mathrm{Ca}$ in SLy5t.
The proton transfer is more subtle, as the nuclear part lowers the barrier, while the Coulomb part increases it~\cite{Godbey2017_PRC95-011601}.
Another effect, the couplings to the low-lying vibration states in $^{48}\mathrm{Ca}$ and $^{56}\mathrm{Ni}$,
increases the dynamical barrier. The two dynamical effects, including the modifications of vibration states and nucleon transfer,
are opposite in fusion dynamics, and hence cancel each other to some extent.
The lower dynamical barrier in SLy5t as seen in Fig.~\ref{Fig:barriers}
indicates the nucleon transfer plays a dominant role in $^{48}\mathrm{Ca}$+$^{56}\mathrm{Ni}$.

\begin{figure}
\centering
\includegraphics[width=8 cm]{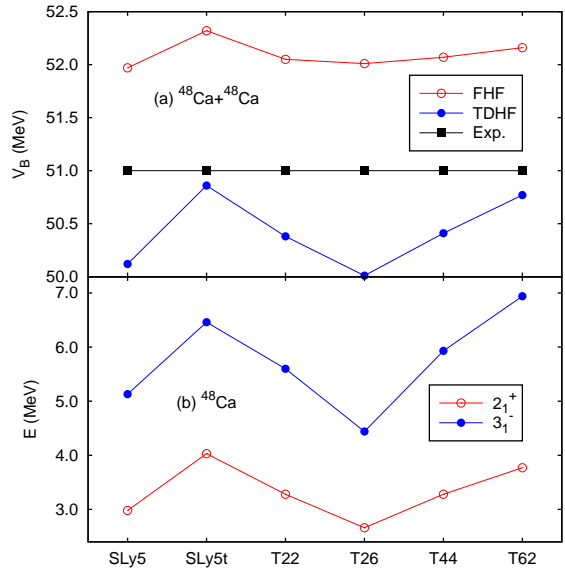}
\caption{(Color online.) Static FHF and dynamic TDHF fusion barriers with six Skyrme forces as well as
experimental value~\cite{Stefanini2009_PLB679-95} for $^{48}\mathrm{Ca}+\mathrm{^{48}Ca}$ in the upper panel, and the peak energies
of $2_1^+$ and $3_1^-$ states in the lower panel.}
\label{Fig:Ca48Ca48_barrier}
\end{figure}

We now proceed to a comparison among the results of various forces, for which the coupling constants
are listed in Tab.~\ref{tab:CC}, for the reaction $^{48}\mathrm{Ca}+\mathrm{^{48}Ca}$ as an example.
The upper panel of Fig.~\ref{Fig:Ca48Ca48_barrier} shows that the static barrier systematically overestimates the experimental data~\cite{Stefanini2009_PLB679-95} and dynamical barrier,
because the couplings to the internal degrees of freedom have been neglected in static FHF calculations.
The dynamical barrier by the inclusion of tensor force in SLy5t are in better agreement than SLy5.
Note that the improvement is rather remarkable given the fact that no free parameters are adjusted to reproduce the
dynamical properties, e.g., fusion barrier and cross section in TDHF calculations.
However, one should note that this improvement could be spurious as it is partly due to the pushing up
of $E_{3_1^-}$ in $^{48}\mathrm{Ca}$ which disagrees with experimental data. Nevertheless, this shows that the
tensor force plays a significant role on fusion, which is the main focus of this work. A better reproduction
of experimental data requires an improvement of the fitting procedure of the tensor terms,
which is beyond the scope of this work.
For T$22$ and T$44$ the dynamical barriers are similar, indicating the isoscalar tensor coupling has
negligible effect in this reaction. By comparing the results with T$26$, T$44$, and T$62$, the barrier height increases as the isovector tensor
coupling decreases. This clear dependence of isoscalar and isovector tensor coupling may be due to the interplay between tensor terms
and rearrangement of mean-field. Although the isoscalar tensor with the proton and neutron single particle spectrum moving
in the same way could affect these vibration modes, the effect seems to be canceled by the refitting of the parameters.
However, the refitting does not absorb the variations of isovector tensor in the same way.
Another interesting observation is, as shown in the lower panel of Fig.~\ref{Fig:Ca48Ca48_barrier}, the peak energies of $2_1^+$ and $3_1^-$ states
in $^{48}\mathrm{Ca}$ display a very similar dependence on the tensor coupling as the dynamical TDHF barrier behaves. This indicates
the fusion dynamics in symmetric reactions is dominated by the low-lying vibrations of colliding partners.
Due to the subtle interplay between nuclear structure and reaction dynamics,
these T$IJ$ forces may have distinct impact on the different heavy-ion collisions.

\begin{table}
\caption{Isoscalar and isovector spin-current coupling constants in units of MeVfm$^5$.}
\label{tab:CC}
\begin{center}
\begin{tabular*}{0.4\textwidth}{@{\extracolsep{\fill}}rrr}
\hline\hline
Force & $\mathrm{C}^\mathrm{J}_0$ & $\mathrm{C}^\mathrm{J}_1$  \\
\hline
T22   &     0   &      0   \\
T26   &   120   &    120   \\
T44   &   120   &      0   \\
T62   &   120   &   -120   \\
SLy5  &  15.65  &    64.55  \\
SLy5t & -19.35  &   -70.45  \\
\hline \hline
\end{tabular*}
\end{center}
\end{table}

To investigate the importance of tensor terms on dynamical effects such as dissipation, we computed above-barrier fusion cross sections
in $\mathrm{^{40}Ca}+\mathrm{^{40}Ca}$ and $^{48}\mathrm{Ca}+\mathrm{^{48}Ca}$.
For the spin-saturated reaction $\mathrm{^{40}Ca}+\mathrm{^{40}Ca}$, the fusion cross sections with and without tensor force are quite close to each other, which is consistent with the observation of the same fusion barriers between SLy5
and SLy5t as shown in Fig.~\ref{Fig:barriers}. We observe the calculated cross sections overestimate the experimental data ~\cite{Montagnoli2012_PRC85-024607} by about 20\%.
Similar overestimations have been obtained for other systems like $\mathrm{^{16}O}+\mathrm{^{208}Pb}$~\cite{Simenel2008_IJMPE17-31,Simenel2013_PRC88-024617}.

\begin{figure}
\includegraphics[width=8 cm]{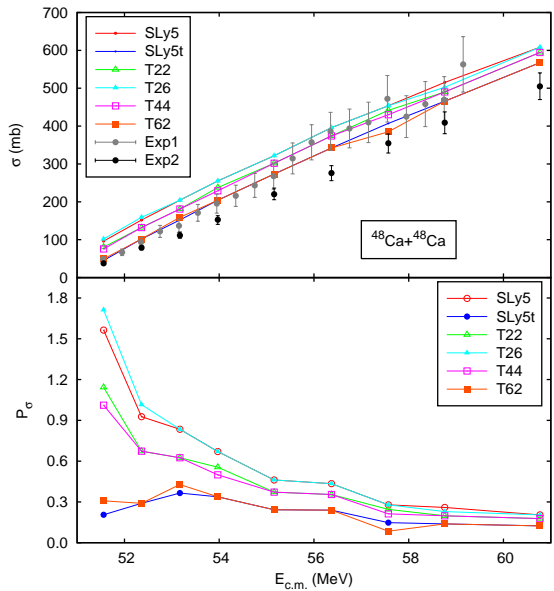}
\caption{(Color online.) The fusion cross sections calculated with various forces as well as the older (denoted as Exp1)
~\cite{Trotta2001_PRC65-011601} and recent experimental data (Exp2)~\cite{Stefanini2009_PLB679-95} for $\mathrm{^{48}Ca}+\mathrm{^{48}Ca}$ in the upper panel,
and the deviation of theoretical fusion cross section from the recent experiment~\cite{Stefanini2009_PLB679-95}, defined as $\mathrm{P_\sigma}=({\sigma_{\mathrm{th}}-\sigma_{\mathrm{exp}}})/{\sigma_{\mathrm{exp}}}$ in the lower panel.}
\label{Fig:fusionca48ca48}
\end{figure}

The fusion cross sections in $^{48}$Ca$+^{48}$Ca with various forces together with the experimental
data~\cite{Trotta2001_PRC65-011601,Stefanini2009_PLB679-95} are given in the upper panel of Fig.~\ref{Fig:fusionca48ca48}.
The older experimental data (denoted as Exp1)~\cite{Trotta2001_PRC65-011601} is systematically higher than the recent data
(Exp2)~\cite{Stefanini2009_PLB679-95}.
The variation between the calculations is globally within the fluctuations of experimental data.
To discriminate the variation between the calculations, the deviation $P_\sigma$ is shown in the lower panel of Fig.~\ref{Fig:fusionca48ca48}.
We see that some forces lead to results in better agreement than others. In particular, the results in SLy5t lead to a much better agreement
than in SLy5. This observation can be traced back to the increase of the dynamical barrier in SLy5t as seen in Fig.~\ref{Fig:Ca48Ca48_barrier}(a).
The T$62$ interaction also leads to very similar cross sections as SLy5t, which is again interpreted by
the fact that the barriers with T$62$ and SLy5t are very close (see Fig.~\ref{Fig:Ca48Ca48_barrier}(a)).
Another interesting observation is that the theoretical cross sections above barrier have all very similar slopes, whatever the forces.
This is a strong indication that dissipation is not significantly affected by the tensor force at the near-barrier energies,
and the effect on the fusion cross section mainly comes from the change of fusion barrier height.

In summary, we incorporate the full tensor force into  TDHF calculations so that both exotic and stable collision partners, as well as their
dynamics, can be described microscopically.
We systematically investigate the role of tensor force on fusion dynamics, and a notable effect is observed in the spin-unsaturated reactions.
The isoscalar and isovector tensor terms in T$IJ$ forces play distinct role
in the fusion dynamics due to the subtle interplay between nuclear structure and reaction dynamics.
Since the near-barrier fusion can be strongly affected by the couplings to intrinsic degrees of
freedom of the collision partners, three dynamical origins, including the vibration states, nucleon transfer,
and dynamical dissipation, are investigated to interpret the observed effects of tensor force in fusion dynamics.
In the symmetric collisions, the pushing-up of low-lying vibration states
is responsible for the observed fusion hindrance. In the asymmetric collisions, the nucleon transfer
and modification of low-lying vibration states may play a similar or opposite effect on the fusion dynamics, depending on
the underlying shell structure of collision partners. The above-barrier fusion cross sections do not show significant effect
of the tensor force on the dynamical dissipation.

\section{Acknowledgments}
This work is partly supported by NSF of China (Grants No. 11175252 and 11575189),
Presidential Fund of UCAS, NSFC-JSPS International Cooperation Program (Grant No. 11711540016),
and by the Australian Research Council Grants No. FT120100760 and DP180100497.
The computations in present work were performed on the High-performance Computing Clusters of SKLTP/ITP-CAS
and Tianhe-1A supercomputer of the CNSC in Tianjin. Part of the calculations have been performed on the NCI National
Facility in Canberra, Australia, which is supported by the Australian Commonwealth Government.

\bibliographystyle{apsrev4-1}
\bibliography{ref}
\end{document}